\documentclass[a4paper,twocolumn,english,pre,nofootinbib]{revtex4-1}
\usepackage[T1]{fontenc}
\usepackage[latin9]{inputenc}
\usepackage{color}
\usepackage{babel}
\usepackage{mathtools}
\usepackage{amsmath}
\usepackage{amssymb}
\usepackage{graphicx}

\makeatletter

\newcommand{\av}[1]{\left\langle {#1} \right\rangle}

\makeatother

\begin{document}
\title{Epidemic Spreading and Aging in Temporal Networks with Memory}
\author{Michele Tizzani}
\affiliation{Dipartimento di Scienze Matematiche, Fisiche e Informatiche, Universit�
degli Studi di Parma, Parco Area delle Scienze, 7/A 43124 Parma, Italy}
\affiliation{INFN, Gruppo Collegato di Parma, Parco Area delle Scienze 7/A, 43124
Parma, Italy}
\author{Simone Lenti}
\affiliation{Dipartimento di Fisica, \textquotedbl Sapienza'' Universit� di Roma,
P.le A. Moro 2, I-00185 Roma, Italy}
\author{Enrico Ubaldi}
\affiliation{ISI Foundation, 10126 Torino, Italy}
\author{Alessandro Vezzani}
\affiliation{Istituto dei Materiali per l'Elettronica ed il Magnetismo (IMEM-CNR),
Parco Area delle Scienze, 37/A-43124 Parma, Italy}
\affiliation{Dipartimento di Scienze Matematiche, Fisiche e Informatiche, Universit�
degli Studi di Parma, Parco Area delle Scienze, 7/A 43124 Parma, Italy}
\author{Claudio Castellano}
\affiliation{Istituto dei Sistemi Complessi (ISC-CNR), via dei Taurini 19, I-00185
Roma, Italy}
\affiliation{Dipartimento di Fisica, \textquotedbl Sapienza'' Universit� di Roma,
P.le A. Moro 2, I-00185 Roma, Italy}
\author{Raffaella Burioni}
\affiliation{Dipartimento di Scienze Matematiche, Fisiche e Informatiche, Universit�
degli Studi di Parma, Parco Area delle Scienze, 7/A 43124 Parma, Italy}
\affiliation{INFN, Gruppo Collegato di Parma, Parco Area delle Scienze 7/A, 43124
Parma, Italy}
\begin{abstract}
Time-varying network topologies can deeply influence dynamical processes
mediated by them. Memory effects in the pattern of interactions among
individuals are also known to affect how diffusive and spreading phenomena
take place. In this paper we analyze the combined effect of these
two ingredients on epidemic dynamics on networks. We study the susceptible-infected-susceptible
(SIS) and the susceptible-infected-removed (SIR) models on the recently
introduced activity-driven networks with memory. By means of an activity-based
mean-field approach we derive, in the long time limit, analytical
predictions for the epidemic threshold as a function of the parameters
describing the distribution of activities and the strength of the
memory effects. Our results show that memory reduces the threshold,
which is the same for SIS and SIR dynamics, therefore favouring epidemic
spreading. The theoretical approach perfectly agrees with numerical
simulations in the long time asymptotic regime. Strong aging effects
are present in the preasymptotic regime and the epidemic threshold
is deeply affected by the starting time of the epidemics. We discuss
in detail the origin of the model-dependent preasymptotic corrections,
whose understanding could potentially allow for epidemic control on
correlated temporal networks. 
\end{abstract}
\date{\today}

\maketitle

\section{Introduction}

In many social and information systems, the time scales for the evolution
of the interaction network are often comparable to the time scales
of the dynamical processes taking place on top of them~\citep{Holme2012}.
The diffusion of online information or the spreading of transmitted
diseases in a population are typical examples of such processes, for
which a focus on a static representation of the network is not able
to capture the very influence of the rapidly varying topology~\cite{Balcan:2009aa,Barrat:2008aa,Eubank:2004aa,PhysRevE.73.046138,bajardi2011,Valdano18}.
Besides, recent advances in technology have allowed to measure and
monitor the evolution of interactions with an unprecedented time resolution~\cite{10.1371/journal.pone.0011596},
calling for new theories to understand the effect of time-varying
topologies on dynamical processes.

Interactions and the creation of links are generated by the agents
activity, a quantity that can be easily measured from available large
scale and time resolved datasets~\cite{Saramaki:2015aa}. An interesting
line of modelling has developed, aiming at including explicitly the
effect of activity distributions on network dynamics: activity-driven
networks~\cite{Perra2012}. In activity-driven models, each agent
is endowed with a degree of freedom that encodes the propensity of
the individual to engage in a social event, establishing a link with
another agent in the system. Notably, measured activities are typically
highly heterogeneous and this has strong effects on network evolution.

When links are randomly established among agents, activity-driven
models have been studied in detail~\cite{Perra:2012fk,Perra2012,PhysRevE.89.032807,Rizzo16},
uncovering the effects of heterogeneous activity distributions on
network topology and on dynamical processes, such as random walks
and epidemic processes.

However, in general agents do not connect randomly to their peers~\cite{Granovetter1977,Miritello2013,Saramaki2014}.
During their activity, individuals remember their friends and their
social circles and they are more inclined to interact with already
known pals, establishing strong and weak ties with their peers~\cite{Dunbar:1992aa,Stiller:2007aa}.
Recent works have tackled this problem by applying a data-driven approach.
A tie allocation mechanism in real systems has been measured, introducing
a memory process on top of activity-driven models~\cite{Karsai:2014aa,Kim2015}.
As reasonably expected, social interactions are not randomly established
but they are rather concentrated towards already contacted nodes,
with a reinforcement process encoded in a single measurable memory
parameter. The memory process tunes the network evolution, that can
be predicted at large times~\cite{Ubaldi:2016aa,Ubaldi2017b,Kim2018},
and it is also expected to influence dynamical processes. Non-Markovian
dynamics can indeed change the spreading rate in a diffusion process,
slowing it down in some cases and speeding it up in others~\cite{Rosvall2013,Scholtes2014,PhysRevLett.112.118702,Lambiotte15m,Karsai11pre,Karsai12plos,Pozzana17}.
Disparate effects have been shown to occur also in epidemic spreading
on activity-driven networks, where memory can lower or increase the
epidemic threshold in SIS or SIR model, respectively~\cite{Sun2015}.
This happens when the epidemic process and the network evolution start
at the same time. However, the network evolution in the presence of
non-Markovian effects could introduce aging in the process~\cite{PhysRevLett.114.108701},
as usually observed in processes with memory also in other fields~\cite{HenkelPt2},
and this could further influence the spreading dynamics.

In this paper, we analyze in detail SIS and SIR epidemic processes
on activity-driven time-varying networks in the presence of memory.
Introducing an \textit{activity-based mean-field} (ABMF) approximation,
we derive analytically, as a function of the activity distribution
and of the parameter tuning the memory, a prediction for the threshold,
holding both for the SIS and the SIR models. The result shows that
memory overall reinforces the effects of activity fluctuations, leading
to a lower value for the epidemics threshold. We prove that ABMF approximation
is equivalent to an epidemic model defined on an effective static
network, that we also investigate. Numerical simulations evidence
that ABMF approach provides exact results when the epidemics start
after the network has evolved for a long time. In this regime indeed
mean field holds since agents have been connected to a large number
of pairs, while the creation of new links becomes negligible.

We show however that strong aging effects are present in the preasymptotic
regime and the epidemic threshold is deeply affected by the starting
time of the epidemics. In particular, at weak memory the creation
of new links cannot be discarded and this increases the epidemic threshold
according to the memoryless predictions. On the other hand, for strong
memory, at short times the dynamics displays correlations among infection
probabilities of nodes which have already been in contact. The correlations
give rise to backtracking effects that cannot be ignored. In this
case, typically the threshold of the SIS and SIR models are respectively
smaller and larger with respect to the mean-field prediction. We explain
in detail the origin of such deviations, opening new perspectives
for epidemic control of disease and information spreading on temporal
networks with high correlations.

The paper is organized as follows. In Section II we first summarize
the activity-driven model for network topology in the presence of
memory and then define epidemic dynamics on top of it. In the next
Section, after a brief recapitulation of the analytical approach to
epidemic dynamics on memoryless activity-driven networks, we describe
in detail how the approach is modified to deal with networks with
memory, deriving predictions for the epidemic threshold. In Section
IV we compare analytical predictions with numerical results, obtained
by considering both an effective static network and the full time-evolution
of the topology. The final Section presents some concluding remarks
and perspectives for future work.

\section{The model}

\subsection{Activity-Driven Networks with memory}

In activity-driven models~\citep{Perra2012}, each node $v_{i}$
($i=1,\dots,N$) of the graph $G_{t}$ has an activity $a_{i}$ assigned
randomly according to a given distribution $F(a)$. The dynamics occurs
over discrete temporal steps of length $\Delta t$. At each step,
with probability $a_{i}\Delta t$ the vertex $v_{i}$ becomes active
and gets linked to $m$ other vertices. Connections last for a temporal
interval $\Delta t$. At the next time step $t+\Delta t$ all existing
edges are deleted and the procedure is iterated. Note that the activity
$a$ has the dimensions of a probability per unit time. Real data
observations indicate that human interactions are very often characterized
by skewed and long tailed activity distributions~\citep{Perra2012}
so $F(a)$ is typically assumed to be a power-law, $F(a)=Ba^{-(\nu+1)}$
with $\varepsilon\le a_{i}\le A$. Since in our simulations we will
keep the time interval $\Delta t=1$, the upper cutoff is naturally
set to $A=1$.

In order to take into account the tendency of individuals to persist
in their social connections, a ``reinforcement'' mechanism has recently
been introduced in activity-driven models~\citep{Karsai:2014aa,Ubaldi:2016aa}.
The nodes are endowed with a memory of their previous contacts and
they have a propensity to establish contacts preferably with individuals
belonging to their social circle. For an active node $v_{i}$, which
has already contacted $k_{i}(t)$ different nodes at time $t$, this
process is described by assuming that the node connects with a new
node with probability 
\begin{equation}
p[k_{i}(t)]=[1+k_{i}(t)/c_{i}]^{-\beta_{i}},\label{eq:rpb}
\end{equation}
while it establishes a connection with a previously contacted node
with complementary probability $1-p[k_{i}(t)]$. The probability depends
on the degree of the integrated network at time $t$, $k_{i}(t)$,
i.e, the number of nodes that $v_{i}$ has contacted up to time $t$.
We will call $A_{ij}(t)$ the adjacency matrix of this integrated
network. { The parameter $\beta_{i}>0$ tunes the memory process.
For $\beta_{i}\approx0$ the probability $p[k_{i}(t)]\approx1$ weakly
depends on the growing degree $k_{i}(t)$, while, at large $\beta_{i}$,
$p[k_{i}(t)]$ rapidly decays with of $k_{i}(t)$. The constant $c_{i}$
sets an intrinsic value for the number of connections that node $v_{i}$
is able to engage in before memory effects become relevant. Empirical
measures on several datasets~\citep{Ubaldi:2016aa} are compatible
with constant values of $\beta_{i}=\beta$ and $c_{i}=c$, so that
the function $p(\cdot)$ turns out to be independent of $i$. In particular,
the exponent $\beta$ ranges from $\beta\approx0.15$ in the citation
networks, to $\beta\approx0.5$ in Twitter mentions and to $\beta\approx1.2$
in the mobile phone calls. In this paper we study the dependence of
the epidemic threshold on the value of the exponent $\beta$, while
we set $c=1$ and $m=1$, with no loss of generality.}

As shown in Ref.~\cite{Ubaldi:2016aa} the asymptotic form of the
degree distribution for the integrated network can be derived analytically.
In particular, in the regime $1\ll k\ll N$ the degree of nodes of
activity $a$ is narrowly distributed around the average value 
\begin{equation}
\bar{k}(a,t)=C(a)t^{1/(1+\beta)},\label{ka}
\end{equation}
i.e. the degree of each node increases sublinearly in time, with a
prefactor depending on its activity. The prefactor $C(a)$ is determined
by the condition 
\begin{equation}
\frac{C(a)}{1+\beta}=\frac{a}{C^{\beta}(a)}+\int da\frac{F(a)a}{C^{\beta}(a)}.\label{Ca}
\end{equation}
In the memoryless case $\beta=0$, where an active node connects always
with a randomly chosen vertex, Eq.~(\ref{Ca}) gives $C(a)=a+\left\langle a\right\rangle $
recovering the result of~\cite{PhysRevE.87.062807}. Hereafter we
will denote in general with $\langle g\rangle=\int daF(a){g(a)}$
the average of a function of the activity $g(a)$ over the network.

\subsection{The epidemic process}

We now turn to the spreading of infectious diseases on activity-driven
temporal networks with memory. We start by considering the standard
Susceptible-Infected-Susceptible (SIS) model, the simplest description
of a disease not conferring immunity. In the SIS model nodes can be
in two states, either susceptible (S) or infectious (I). An infected
node can turn spontaneously susceptible with rate $\mu$, while an
infected node transmits the infection over an edge to a susceptible
neighbor with rate $\lambda$. The two elementary events are therefore:
\begin{equation}
I+S\overset{\lambda}{\longrightarrow}2I\hspace{2cm}I\overset{\mu}{\longrightarrow}S\label{eq:sis}
\end{equation}

In the Susceptible-Infected-Recovered (SIR) model the disease confers
immunity, so that the nodes can be in three states susceptible (S),
infectious (I) and recovered (R) which are immune to a new infection.
The dynamics is described by the following reaction scheme: 
\begin{equation}
I+S\overset{\lambda}{\longrightarrow}2I\hspace{2cm}I\overset{\mu}{\longrightarrow}R\label{eq:sir-1}
\end{equation}

The epidemic process on activity-driven networks is implemented by
iterating discrete time steps of duration $\Delta t$: 
\begin{itemize}
\item at the beginning of each time step there are $N$ disconnected vertices;
\item with probability $a_{i}\Delta t$ a vertex $v_{i}$ becomes active
and connects to a previously linked node with probability $1-p(k_{i})$,
or with a new node $v_{j}$ with probability $p(k_{i})$, in this
second case $k_{i}(t)$, $k_{j}(t)$ and $A_{ij}(t)$ are increased
by one unit;
\item if one of the nodes connected by the link is infected and the other
one is susceptible, the susceptible becomes infected with probability
$\lambda$;
\item a vertex $v_{j}$, if infected, becomes susceptible (SIS), or recovers
(SIR) with probability $\mu\Delta t$. 
\end{itemize}
In activity-driven models, $\lambda$ is a pure number, i.e. the probability
that in a single contact the infection is actually transmitted, while
$\mu$ is still the rate of recovery for a single individual. Ignoring
the inhomogeneity in the activities, one can estimate the total rate
for the infection process per node as $\lambda\left\langle k'\right\rangle $,
where $\left\langle k'\right\rangle =2\left\langle a\right\rangle $
is the average degree per unit time; this is the quantity to be compared
with the recovery rate per node $\mu$.

\section{Analytical results}

\subsection*{Epidemics on memoryless activity-driven networks}

The epidemic spreading for the memoryless case $\beta=0$ has been
studied in~\citep{Perra2012} adopting an ABMF approach. The epidemic
state of a node, when averaged over all possible dynamical evolutions,
only depends on the value of its activity $a_{i}$. In particular,
one can define the probability $\rho(a_{i},t)$ that a node with activity
$a_{i}$ is infected at time $t$. The corresponding evolution equation
is: 
\begin{equation}
\begin{array}{rcl}
\partial_{t}\rho(a_{i}) & = & -\mu\rho(a_{i})+\lambda[1-\rho(a_{i})]\\
 &  & \frac{1}{N-1}\underset{j\ne i}{\sum}\left[a_{i}\rho(a_{j})+a_{j}\rho(a_{j})\right].
\end{array}\label{ML}
\end{equation}
The first term on the right side is due to recovery events; the second
term takes into account the event that a susceptible node of class
$a_{i}$ becomes active and contracts the disease by connecting to
an infected individual, while the third term is the analogous term
for the case of a susceptible node that, independently of her own
activity, is contacted by an infected active individual.

The description in terms of quantities that only depend on the activity
is conceptually analogous to the heterogeneous-mean-field approach
for dynamical processes on static networks~\cite{PastorSatorras2001}.
In that case, one assumes that the only property determining the epidemic
state of a node is the degree $k$ and then derives equations for
the probabilities $\rho_{k}$. An important difference must however
be stressed. Assuming the epidemic state to depend only on the degree
is an approximation for static networks, because it neglects the quenched
nature of the network structure that makes properties of nodes, with
the same degree but embedded in different local environments, different.
In practice, this assumption is equivalent to replacing the actual
adjacency matrix of the network ($A_{ij}$ equal to 0 or 1 depending
on the presence of the connection between $v_{i}$ and $v_{j}$) with
an annealed adjacency matrix ${P}_{ij}=k_{i}k_{j}/(\left\langle k\right\rangle N)$~\cite{Dorogovtsev2008},
expressing the probability that vertices $v_{i}$ and $v_{j}$ with
degree $k_{i}$ and $k_{j}$ are connected. The annealed approach
is an approximation for static networks, while it is exact for networks
where connections are continuously reshuffled at each time step of
the dynamics, since the reshuffling process destroys local correlations.
Since in memoryless activity-driven networks connections are extracted
anew at each time step, the ABMF approach provides exact results in
this case.

Equation~(\ref{ML}) can be analyzed by means of a linear stability
analysis, yielding, for large $N$, the threshold~\cite{Perra2012}
\begin{equation}
\left(\frac{\lambda}{\mu}\right)_{{\rm ML}}=\frac{1}{\left\langle a\right\rangle +\sqrt{\left\langle a^{2}\right\rangle }}.\label{eq:eth}
\end{equation}
The same result can be derived for the SIR case.

\subsection*{Epidemics on activity-driven networks with memory}

\subsubsection*{Individual-based mean-field approach}

In presence of memory, interactions occur preferably with a subset
of the other nodes (the social circle) creating correlations. Therefore,
we implement a different, individual-based, mean-field approach, keeping
explicitly track of the evolution of social contacts (i.e. of the
memory). Let us first consider the SIS model. The observable of interest
is the probability $\rho_{i}(t)$ that node $v_{i}$ is infected at
time $t$. Its evolution can be written as 
\begin{widetext}
\begin{equation}
\begin{array}{rcl}
\partial_{t}\rho_{i}(t)=-\mu\rho_{i}(t)+\lambda\left[1-\rho_{i}(t)\right] & \!\!\!\! & \left\{ \sum_{j}a_{i}\left[1-p(k_{i})\right]\frac{A_{ij}(t)}{k_{i}}\rho_{j}(t)+\underset{j\nsim i}{\sum}a_{i}p(k_{i})\frac{1}{N-k_{i}-1}\rho_{j}(t)\right.+\\
 & \!\!\!\! & \left.\sum_{j}a_{j}\left[1-p(k_{j})\right]\frac{A_{ij}(t)}{k_{j}}\rho_{j}(t)+\underset{j\nsim i}{\sum}a_{j}p(k_{j})\frac{1}{N-k_{j}-1}\rho_{j}(t)\right\} 
\end{array}\label{eq:ibmf}
\end{equation}
\end{widetext}

Here $j\nsim i$ indicates the sum over the nodes $j$ not yet connected
to $i$, $N-k_{j}(t)-1$ is their number. The quantity $A_{ij}(t)$
is the adjacency matrix of the time-integrated network at time $t$,
i.e., it is equal to 1 if $v_{i}$ and $v_{j}$ have been in contact
at least once in the past and 0 otherwise. In Eq. (\ref{eq:ibmf}),
the only approximation made is that the dynamical state of every node
is considered to be independent of the state of the partner in the
interaction; in other words, we neglect the existence of dynamical
correlations among nodes, which are created by the partially quenched
nature of the interaction pattern due to memory. It is exactly the
same approximation that is involved by the individual-based mean-field
approach for static networks~\cite{PastorSatorras15}.\\

The first term on the right hand side of Eq.~(\ref{eq:ibmf}) is
the recovery rate of $\rho_{i}(t)$. The second term, describing the
infection process, is the product of $\lambda$ times the probability
for $v_{i}$ to be susceptible and, in curly brackets, the fraction
of infected nodes contacted by $v_{i}$ per unit time. In the curly
brackets, the first and the second term describe the case where $v_{i}$
is active and connects to the infected node $v_{j}$ taking into account
that the link can be an old or a new one respectively. In the same
way, the third and the fourth term represent the probabilities that
$v_{i}$ is contacted by an infected and active node $v_{j}$.

Since both $A_{ij}(t)$ and $k_{i}(t)$ depend on the evolution time
$t$, the behavior of the epidemics can strongly depend on the starting
time of the outbreak, giving rise to aging effects that will be investigated
in numerical simulations. When the epidemic starts at very large times,
an analytic approach can be considered. In this regime, with $1\ll k_{i}(t)\ll N$,
we expect that the creation of new contacts can be ignored and that
the dynamical correlations are asymptotically negligible, since the
connectivity of the integrated network becomes large. If the epidemic
starts at very large times, therefore, we can apply an heterogeneous
mean-field approximation for $A_{ij}(t)$, allowing for an analytical
solution of the problem which we expect to be asymptotically exact.

\subsubsection*{The behavior for large times}

Let us consider the regime of large times, where $1\ll k_{i}(t)\ll N$
for all nodes. This means that each node has already had a large number
of contacts, but that number is not too large, so that the integrated
network cannot be considered as a complete graph, i.e., it is still
sparse. In the limit of large $N$ there is clearly a large temporal
interval such that this condition is fulfilled. The condition $1\ll k_{i}(t)\ll N$
allows us to replace in Eq.~(\ref{eq:ibmf}) $N-k_{i}(t)-1$ with
$N$ and $p(k_{i})$ with $(k_{i}(t))^{-\beta}$. Considering only
leading terms Eq.~(\ref{eq:ibmf}) becomes 
\begin{equation}
\partial_{t}\rho_{i}(t)=-\mu\rho_{i}(t)+\lambda\left[1-\rho_{i}(t)\right]\sum_{j}A_{ij}(t)\left(\frac{a_{i}}{k_{i}}+\frac{a_{j}}{k_{j}}\right)\rho_{j}(t).\label{ibmf2}
\end{equation}

\subsubsection*{The linking probability}

To proceed further we perform the equivalent of the heterogeneous
mean-field approximation for static networks, i.e., we replace the
time-integrated adjacency matrix $A_{ij}(t)$ with its annealed form,
$P_{ij}(t)$, i.e., the probability that $v_{i}$ and $v_{j}$ have
been in contact in the past. The evolution of $P_{ij}(t)$ is described
by the master equation: 
\begin{equation}
\partial_{t}P_{ij}(t)=\left[\frac{a_{i}p(k_{i})}{N-k_{i}-1}+\frac{a_{j}p_{j}(k_{j})}{N-k_{j}-1}\right]\left[1-P_{ij}(t)\right].\label{Pij}
\end{equation}
In Eq. (\ref{Pij}) $P_{ij}$ grows either because the node $v_{i}$
activates (probability per unit time $a_{i}$), it creates a new connection
{[}probability $p(k_{i})${]} and the new partner is $v_{i}$ {[}probability
$(N-k_{i}-1)^{-1}${]} or because of the event with the role of $v_{i}$
and $v_{j}$ interchanged.

In the temporal interval of interest we can use again the relations
holding for large times $p(k_{i})\approx k_{i}^{-\beta}$ and $N-k_{j}-1\approx N$.
Moreover, for large times, the degree of a node of activity $a_{i}$
can be estimated by its average value $\bar{k}(a_{i},t)$, given by
Eq. (\ref{ka}). So we obtain 
\begin{equation}
\partial_{t}P_{ij}(t)=[1-P_{ij}(t)]\frac{g(a_{i})+g(a_{j})}{Nt^{\frac{\beta}{1+\beta}}},\label{Pij2}
\end{equation}
where we have defined 
\begin{align}
g(a_{i}) & =a_{i}/[C(a_{i})]^{\beta}.\label{eq:ga}
\end{align}
Eq.~(\ref{Pij2}) can be readily solved, yielding 
\begin{equation}
P_{ij}(t)=1-e^{-\frac{(1+\beta)t^{1/(1+\beta)}}{N}[g(a_{i})+g(a_{j})]}
\end{equation}
In the regime $t^{1/(1+\beta)}\ll N$, $P_{ij}(t)$ becomes 
\begin{equation}
P_{ij}(t)=(1+\beta)\frac{t^{1/(1+\beta)}}{N}[g(a_{i})+g(a_{j})].\label{LP}
\end{equation}
Notice that $P_{ij}(t)$ is a topological feature of the activity-driven
network, independent of the epidemic process.

\subsubsection*{Asymptotic ABMF equation}

We now introduce into Eq.~(\ref{ibmf2}) the annealed expression
for the integrated adjacency matrix $A_{ij}(t)\approx P_{ij}(t)=P(a_{i},a_{j},t)$
and for the connectivity $k_{i}(t)=\bar{k}(a_{i},t)$. In this way
the equations depend on the nodes $v_{i}$ and $v_{j}$ only through
their activities $a_{i}$ and $a_{j}$. The equation for the probability
$\rho(a,t)$ that a generic node of activity $a$ is infected at time
$t$ is therefore: 
\begin{widetext}
\begin{equation}
\begin{array}{rcl}
\partial_{t}\rho(a,t)=-\mu\rho(a,t)+\lambda\left[1-\rho(a,t)\right] & \!\!\!\! & \left\{ \frac{ag(a)}{g(a)+\av{g}}\int da'F(a')\rho(a',t)+\frac{a}{g(a)+\av{g}}\int da'F(a')\rho(a',t)g(a')+\right.\\
 & \!\!\!\! & g(a)\int da'F(a')\frac{a'}{\left(g(a')+\av{g}\right)}\rho(a',t)+\left.\int da'F(a')\frac{a'g(a')}{\left(g(a')+\av{g}\right)}\rho(a',t)\right\} 
\end{array}\label{eq:cti}
\end{equation}
\end{widetext}

where we have replaced the sums over nodes with integrals over the
activities $1/N\sum_{j}\to\int da'F(a')$ and used Eq.~(\ref{Ca}),
which can be rewritten as 
\begin{equation}
C(a)=(1+\beta)\left[g(a)+\av{g}\right].\label{eq:ca-1}
\end{equation}

Eq.~(\ref{eq:cti}) is effectively an ABMF approach, since all the
information on the behavior of the node $v_{i}$ depends on its activity
$a_{i}$. Note that, although Eqs.~(\ref{eq:ibmf}) and~(\ref{ibmf2})
described the dynamics of the individual node, the further approximation
underlying Eq.~(\ref{Pij}) has transformed the approach into an
ABMF one, conceptually analogous to the heterogeneous mean-field approximation
on static networks, where all the information on node $v_{i}$ is
encoded in its degree $k_{i}$.

It is important to remark that in Eq.~(\ref{ibmf2}) the time dependencies
of $P(a_{i},a_{j},t)\propto t^{1/(1+\beta)}$ and of the average degree
$\bar{k}(a_{i},t)\propto t^{1/(1+\beta)}$ cancel out, so that the
right hand side of Eq.~(\ref{eq:cti}) does not depend explicitly
on time. This suggests that in this temporal regime the epidemic can
be seen as an activity-driven process taking place on an effective
static graph, where the probability for nodes $v_{i}$ and $v_{j}$
to be linked is given by Eq.~(\ref{LP}) and the quantity $t^{1/(1+\beta)}/N$
is a fixed quantity $\tau$ whose value only determines the average
degree of the network. Performing simulations over an ensemble of
these effective static networks and averaging the results one should
then reproduce the predictions of the ABMF approach, Eq.~(\ref{eq:cti}).

From Equation~(\ref{eq:cti}), by performing a linear stability analysis
around the absorbing state $\rho(a,t)=0$ (see Appendix), it is possible
to compute analytically the epidemic threshold $(\lambda/\mu)_{c}$,
for any value of the reinforcement parameter $\beta$ and of the exponent
of the analytical distribution $\nu$. Since for large times the node
degrees diverge and correlations can be neglected, we expect the linear
stability analysis to provide the correct estimate of the epidemic
threshold when the epidemics start at very long times i.e. when the
degrees $k_{i}(t)$ have already become very large.

\begin{figure}
\includegraphics[width=9cm,height=6cm]{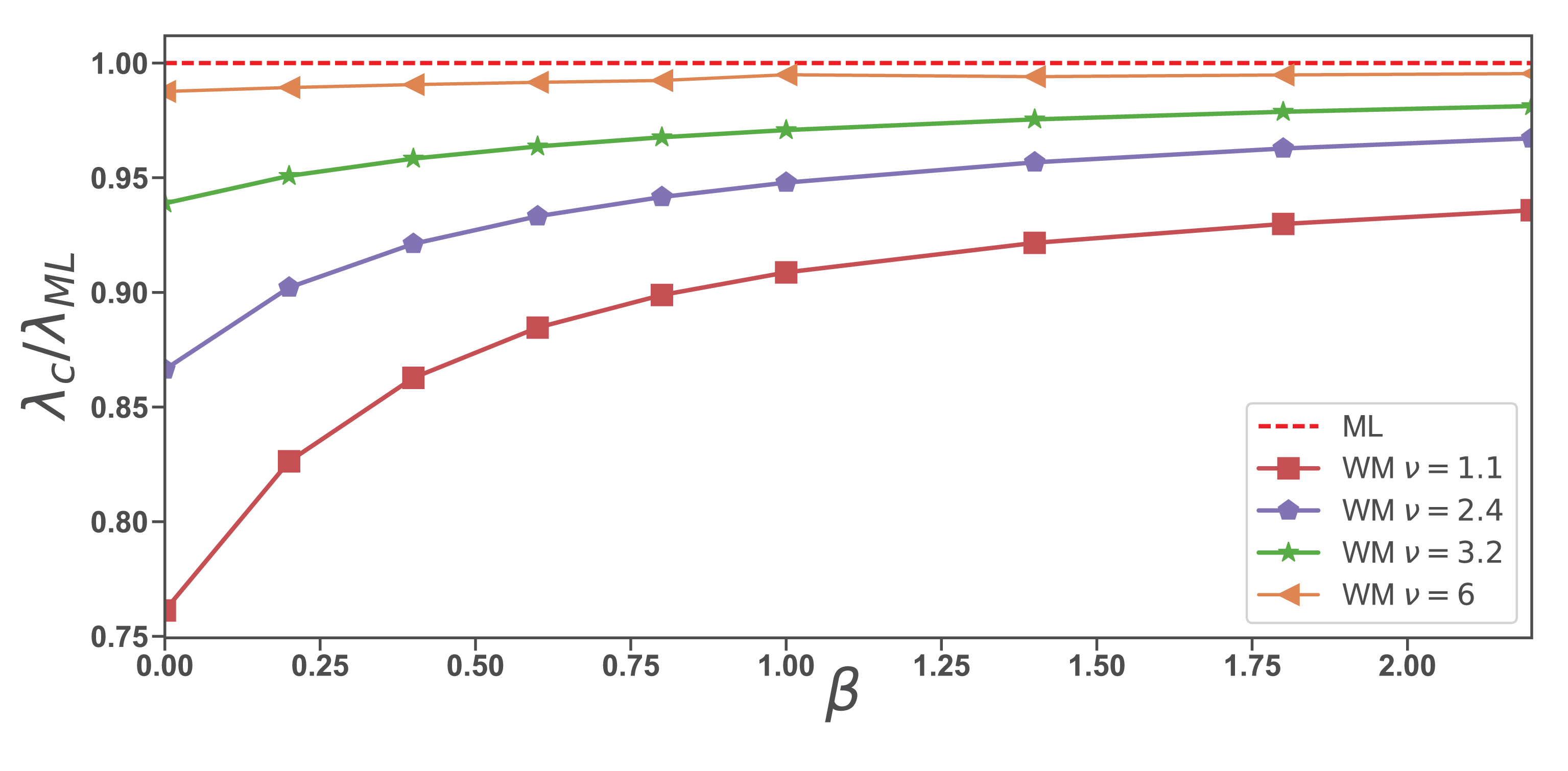} \caption{\label{fig:fig4}Plot of the ratio $\lambda_{c}/\lambda_{ML}$ between
the epidemic thresholds in the memory (WM) and in the memoryless (ML)
cases, for different values of the exponent $\nu$ of the distribution
$F(a)=Ba^{-(\nu+1)}$. The dashed line is the mean-field memoryless
results, while the solid lines are the outcomes of the ABMF equations
in presence of memory. }
\end{figure}

The results of the linear stability analysis are presented in Fig.~\ref{fig:fig4}
showing that the thresholds are smaller than in the memoryless case.
This lower value is a consequence of the fact that memory reinforces
the activity fluctuations, and in these models fluctuations clearly
reduce the the epidemic threshold, as shown by Eq.~(\ref{eq:eth}).
The effect can be simply understood since nodes with large activity
have also a large degree, therefore they are easily involved in epidemic
contacts not only because they are frequently activated but also because
they are frequently contacted by other nodes. In this way memory reinforces
the effect of activity fluctuations. In this framework, Fig.~\ref{fig:fig4}
also shows that at large $\nu$ i.e. for increasingly smaller fluctuations,
the difference with the memoryless model vanishes. In particular,
for $F(a)=\delta(a-a_{0})$ i.e. when the activity does not fluctuate,
one obtains from Eq.(\ref{eq:cti}) $\partial_{t}\rho(t)=-\mu\rho(t)+2a_{0}\lambda[1-\rho(t)]$
that is the same equation of the memoryless case. This also explains
the quite surprising observation that the threshold is a growing function
of $\beta$, converging to the memoryless case as $\beta\to\infty$.
Indeed, the tail of the degree distribution decays at large $k$ as
$k^{-[(1+\beta)\nu+1]}$ ~\citep{Ubaldi:2016aa}, therefore at large
$\beta$ we get a faster decay and smaller degree fluctuations. For
the same reason, in the limit $\beta\to0$ the difference with the
memoryless case is maximal, since degree inhomogeneities are stronger
in this case.

We remark that in Eq.~(\ref{eq:cti}), as in the memoryless case,
dynamical correlations are ignored. However, we expect that at finite
times, due to the finite connectivity of the integrated graph, the
effect of correlations becomes important. The memory process leads
to the formation of small clusters of mutually connected high activity
vertices, which become reservoirs of the disease in the SIS model.
The high frequency of mutual contacts allows for reinfection, favouring
the overall survival of the epidemic spreading in the system. In this
way, social circles with high activity play a role analogous to that
played by the max K-core or the hub and its immediate neighbors for
SIS epidemics in static networks~\cite{Castellano2012,Castellano2017}.
To clarify the effect of dynamical correlations at finite time, in
the next Section we compare the analytical predictions with results
of numerical simulations. As a final remark, we note that, in the
asymptotic ABMF approach, the linear stability analysis also holds
for the SIR model, implying that the epidemic threshold is the same
of the SIS model. However, in the SIR model reinfection is not allowed
so that the initial presence of small clusters of mutually connected
high activity vertices effectively inhibits the spread of the disease.
For this reason, we expect that finite connectivity (i.e. finite time)
increases the epidemic threshold with respect to the mean-field result,
as we will check in numerical simulations.

\section{Numerical simulations}

\subsection*{SIS model on the effective static network}

As discussed above, Eq.~(\ref{eq:cti}) can be interpreted as a heterogeneous
mean-field equation for a SIS epidemic on an effective static network
where the probability that $v_{i}$ and $v_{j}$ are connected is
\begin{equation}
P_{ij}=P(a_{i},a_{j})=\tau(1+\beta)[g(a_{i})+g(a_{j})].\label{eq:pijs}
\end{equation}
Here $\tau\ll1$ is a constant, $g(a)=a/[C(a)]^{\beta}$ and $C(a)$
is a function that can be evaluated numerically for $\beta>0$, while
for $\beta=0$ it takes the simple form $C(a)=a+\av{a}$. The constant
$\tau$ can be tuned in order to set the average degree of the network,
because 
\begin{equation}
k(a)=N\int da'F(a')P(a,a')=(1+\beta)N\tau[g(a)+\av{g}],
\end{equation}
so that 
\begin{equation}
\av{k}=\int da'F(a')k(a')=2(1+\beta)N\tau\av{g}.
\end{equation}
We now study the SIS epidemic evolution on the effective static network.

Given the activity of each node, extracted according to the distribution
$F(a)$, for each of the possible pairs of nodes we place an edge
with probability given by Eq.~(\ref{eq:pijs}). On top of this quenched
topology we run a memoryless activity-driven SIS dynamics, starting
with 10\% of the nodes in the infected state, until the stationary
state state is reached and we record the fraction of infected nodes.
We repeat the procedure many times for each value of $\lambda$, while
$\mu$ is fixed to 0.015. We determine the threshold as the position
of the maximum of the susceptibility~\cite{Ferreira2012} $\chi=N(\overline{\rho^{2}}-\overline{\rho}^{2})/\overline{\rho}$,
where the overbar denotes the average over dynamical realizations
at fixed topology. We repeat all this for several networks obtained
using different sequences of activities and different samplings of
$P_{ij}$ and we average the epidemic thresholds found for each of
them.

\begin{figure}
\centerline{\includegraphics[width=9cm,height=6cm]{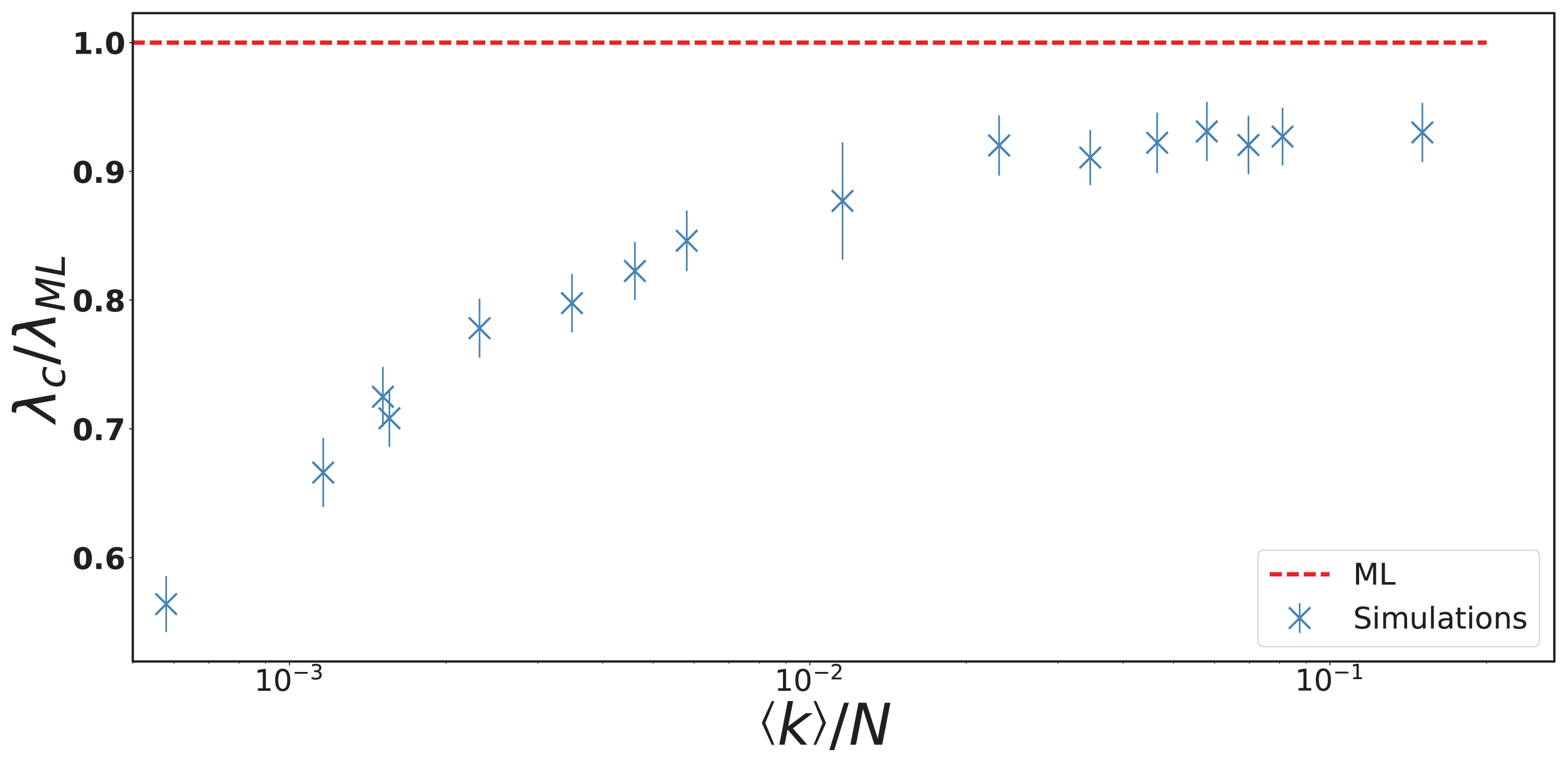}}
\caption{SIS model on the effective static network: $\nu=2.4$, $\epsilon=0.01$,
$\mu=1.5\cdot10^{-2}$ and $N=10^{4}$. Ratio between the epidemic
threshold found in simulations and the result of the memoryless model
in Eq.(\ref{eq:eth}), as a function of $\log(\langle k\rangle/N)$
. For $\langle k\rangle/N>0.01$, we observe practically no dependence
on $\av{k}$.}
\label{indepfromk} 
\end{figure}
We first check that, as long as $1\ll\av{k}\ll N$, the results are
independent of the exact value of $\av{k}$, as predicted by the theory.
Fig.~\ref{indepfromk} shows, for $\beta=1$, that the effective
threshold initially grows with $\av{k}$ but then reaches a plateau,
in accordance with the expectations.

\begin{figure}
\centerline{\includegraphics[width=9cm,height=6cm]{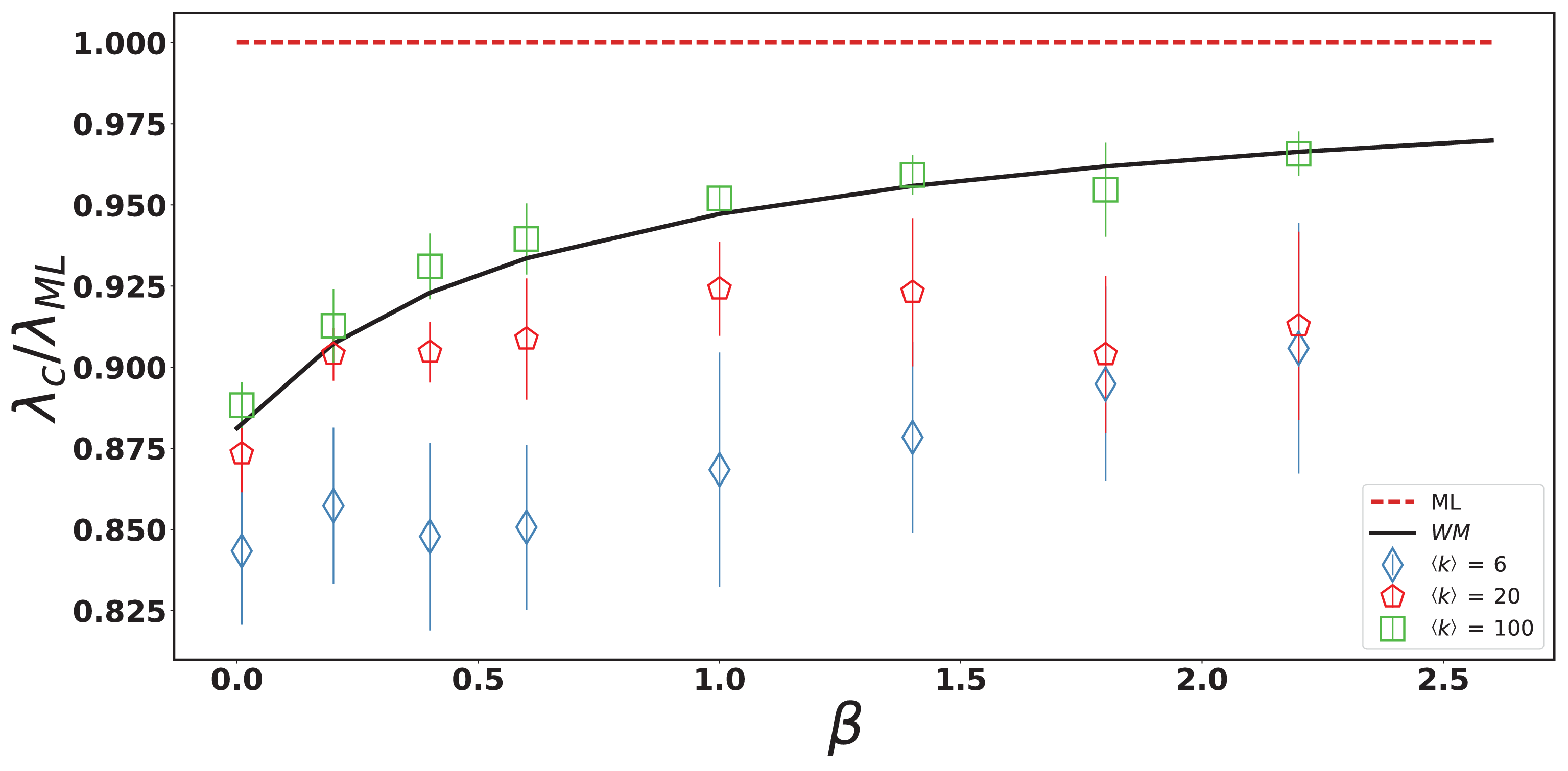}} \caption{SIS model on the effective static network. Ratio between the epidemic
threshold with memory and the epidemic threshold of the memoryless
case as a function of the reinforcement parameter $\beta=[0.01,0.2,0.4,0.6,1,1.4,1.8,2.2]$,
with $\nu=2.4$, $\epsilon=0.01$, $\mu=1.5\cdot10^{-2}$, $N=5\cdot10^{4}$.
The points are averages of different realizations of the network with
different sequences of activity $a_{1},a_{2},\ldots,a_{N}$: 32 realizations
for $\av{k}=6$, 16 realizations for $\av{k}=20$, 4 realizations
for $\av{k}=100$}
\label{staticratio} 
\end{figure}

In Fig.~\ref{staticratio} we report the dependence of the effective
epidemic threshold as a function of $\beta$ for three values of the
average degree $\av{k}$, compared with the predictions of the ABMF
theory with and without memory. { We observe that, as $\av{k}$ grows,
numerical results tend to coincide with theoretical predictions. A
nice agreement with the ABMF results is obtained for $\av{k}\approx100$,
which is a large but realistic number of connections in a social systems.
We remark that $\av{k}\approx100$ is large enough for observing mean
field behavior but also it is much smaller than the total number of
sites $N=5\cdot10^{4}$ so that the system is not fully connected
and degree fluctuations are important.} On the other hand, for small
values of $\av{k}$ the value of the threshold is smaller than the
one predicted theoretically. Indeed, on effective static networks
with small connectivity we expect the presence of clusters of mutually
interconnected nodes to be relevant, as they are able to reinfect
each other several times. It is well known that for the SIS model
these backtracking effects tend to lower the epidemic threshold since
social circles with high activity favor the overall survival of the
epidemic.

\subsection*{Epidemics on time-evolving networks}

Let us now consider simulations of the epidemic spreading on the full
time evolving network. We consider a graph of size $N=5\cdot10^{4}$
with activity distributed according to $F(a)=Ba^{-(\nu+1)}$ ($\nu=2.4$)
and a cutoff $\varepsilon=10^{-2}$. To extract the activities of
individual nodes we perform an importance sampling so that, even in
the finite size system, the moments $\left\langle a\right\rangle $
and $\left\langle a^{2}\right\rangle $ coincide with their expected
values.

We first start the temporal evolution of the network and at a later
time $t_{0}$ we let the epidemic begin. We evaluate at $t_{0}$ the
average connectivity of the nodes $\left\langle k\right\rangle _{0}$
which measures the evolution of the network at the starting time.
In both the SIS and SIR models, we use two different initial conditions.
The first is to randomly select (RA) the node to infect at time $t_{0}$,
while the second is to infect the most active node (MA) at time $t_{0}$.
We keep the recovery rate fixed at $\mu=1.5\cdot10^{-2}$ and vary
the probability of infection $\lambda$ to study the dependence of
its critical value on the memory parameter $\beta$.

\subsubsection*{SIS model}

\begin{figure}
\centerline{\includegraphics[width=9cm,height=6cm]{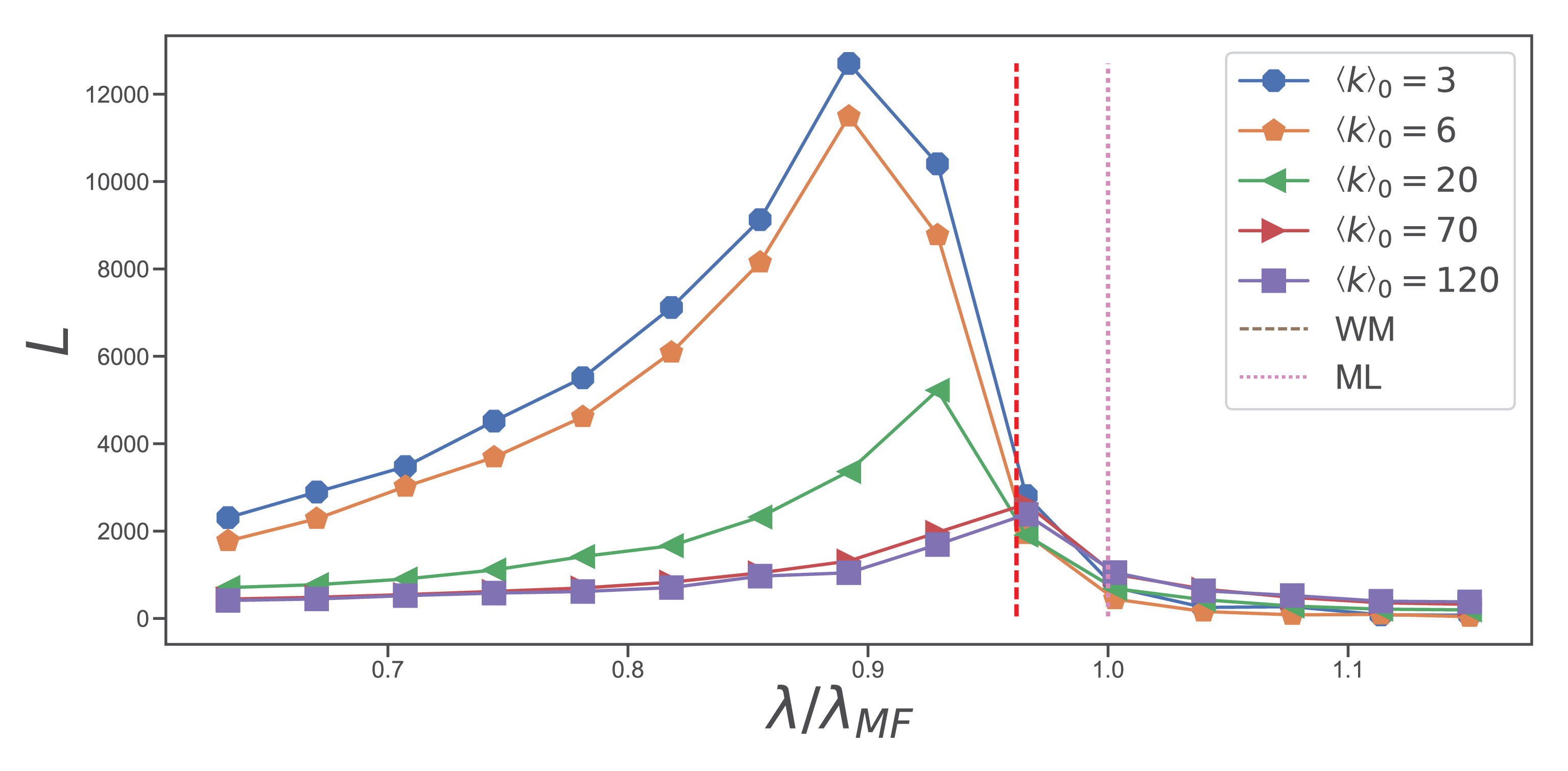}}
\caption{\label{fig:fig1-1}SIS epidemic process on the activity driven network
MA. Lifespan (L) as function of the ratio between the epidemic threshold
with memory and the epidemic threshold of the memoryless for different
values of $\left\langle k\right\rangle _{0}$. $N=5\cdot10^{4}$,
$\nu=2.4$, $a\in[10^{-2},1]$. We consider $4\cdot10^{3}$ epidemic
realizations for each value of $\lambda$. Dynamics starts from the
most active site and at small $\av{k}_{0}$ back-tracking effects
are dominant favouring the epidemic spreading; this on one side lowers
the value of the threshold (value of $\lambda$ corresponding to the
peak) but also increases the lifespan of the system at small $\lambda$.}
\end{figure}

In the SIS model, we determine the epidemic threshold using the lifespan
method~\citep{Boguna2013}. We plot (see Fig. \ref{fig:fig1-1}),
as a function of the parameter $\lambda$, the average lifespan of
simulations ending before the coverage (i.e., the the fraction of
distinct sites ever infected) reaches a preset value that we take
equal to $1/2$. The threshold is estimated as the value of $\lambda$
for which the lifespan has a peak.

\begin{figure}
\centerline{\includegraphics[width=9cm,height=6cm]{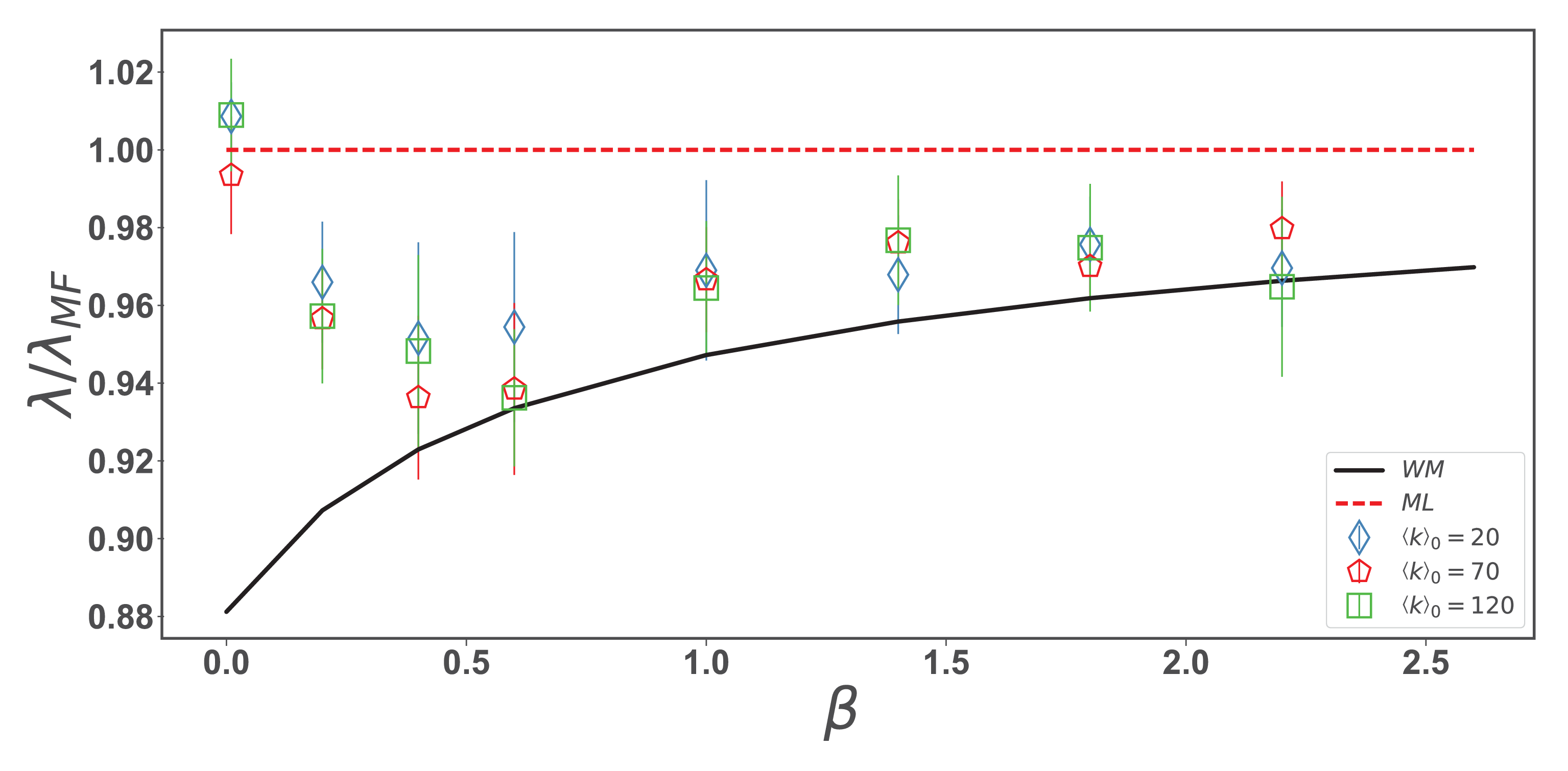}}
\caption{\label{fig:fig1}SIS epidemic process on the activity driven network,
RA. Ratio between the epidemic threshold with memory and the epidemic
threshold of the memoryless case as a function of the reinforcement
parameter $\beta=[0.01,0.2,0.4,0.6,1,1.4,1.8,2.2]$. $N=5\cdot10^{4}$,
$\nu=2.4$, $a\in[10^{-2},1]$. The dotted line represents the memoryless
result (ML), the solid line is the analytical prediction obtained
from Eq.~(\ref{eq:cti}) in the memory case (WM). We consider $4\cdot10^{3}$
epidemic realizations for each value of $\lambda$.}
\end{figure}

\begin{figure}
\centerline{\includegraphics[width=9cm,height=6cm]{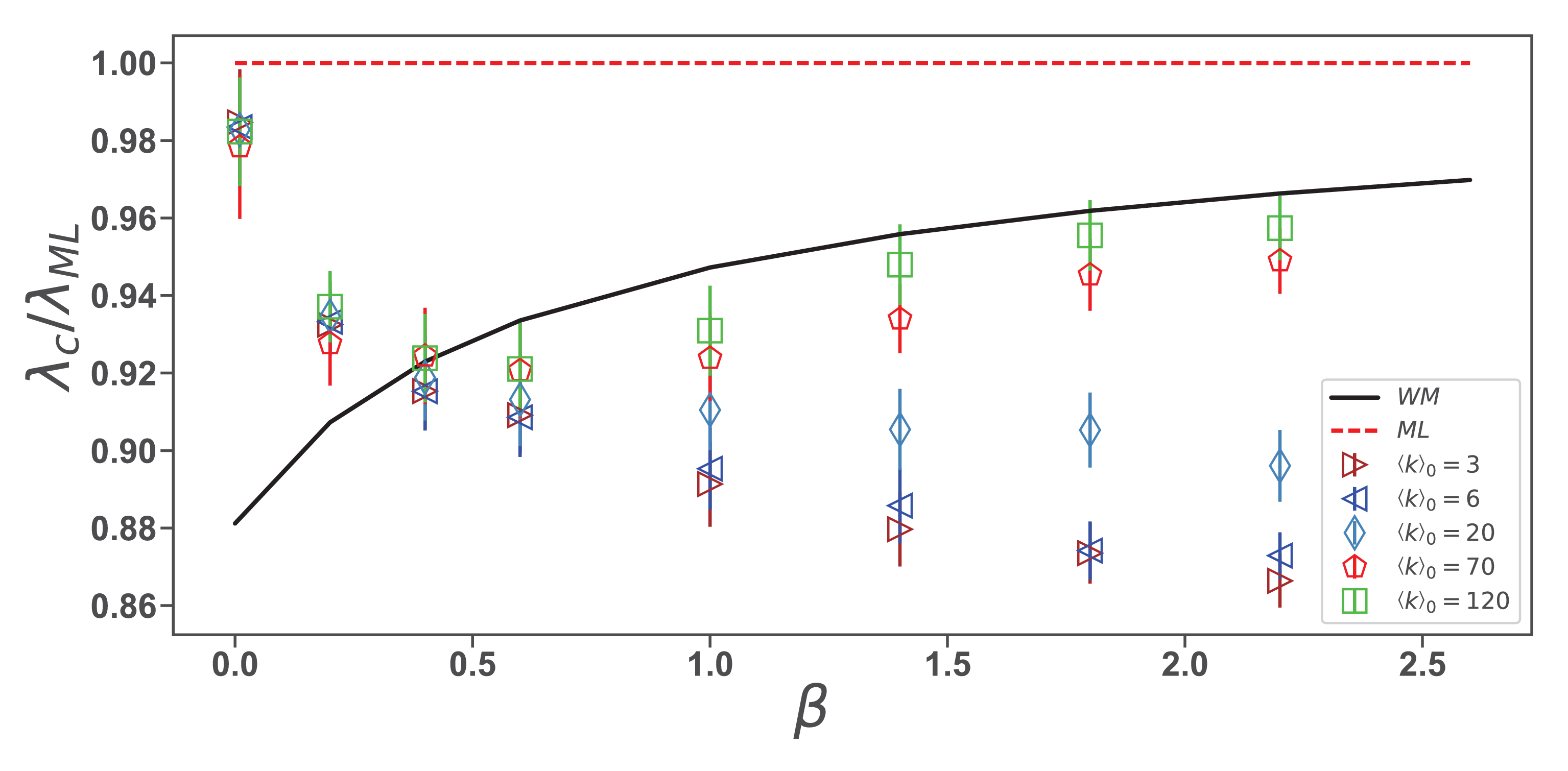}}
\caption{\label{fig:fig2}SIS epidemic process on the activity driven network,
MA. Ratio between the epidemic threshold with memory and the epidemic
threshold of the memoryless case as a function of the reinforcement
parameter $\beta=[0.01,0.2,0.4,0.6,1,1.4,1.8,2.2]$. $N=5\cdot10^{4}$,
$\nu=2.4$, $a\in[10^{-2},1]$. The dotted line represents the memoryless
result (ML), the solid line is the analytical prediction obtained
from Eq.~(\ref{eq:cti}) (WM). We consider $4\cdot10^{3}$ epidemic
realizations for each value of $\lambda$.}
\end{figure}

The epidemic thresholds of numerical simulations are compared with
theoretical predictions in Fig.~\ref{fig:fig1} (RA case) and \ref{fig:fig2}
(MA case). Numerical results converge toward the analytical prediction
as $\left\langle k\right\rangle _{0}$ becomes larger, while there
are strong deviations for small $\left\langle k\right\rangle _{0}$.

{ For small $\left\langle k\right\rangle _{0}$ two competing effects
are at work. First, infections are mediated by an effective static
network with small connectivity, therefore we expect backtracking
effects to enhance epidemic spreading and to lower the threshold.
Second, for small $\left\langle k\right\rangle _{0}$, moves connecting
new partners are also possible. In these moves nodes are chosen randomly
in the whole system according to a memoryless dynamics, which displays
a higher epidemic threshold. So there exists a competition between
backtracking correlations and memoryless moves which reduce and increase
the threshold, respectively. Clearly for large $\left\langle k\right\rangle _{0}$
both effects become negligible and the ABMF result is recovered.

Figs.~\ref{fig:fig1} and~\ref{fig:fig2} show that at large $\beta$
backtracking effects (leading to small thresholds) are strong when
the evolution starts from the most active site, while they are negligible
with random initial conditions. The most active node indeed has the
largest degree and she forms a cluster of highly activated nodes where
the high frequency of mutual contacts allows for reinfections and
positive correlations. Conversely, the average site has a small connectivity
and can activate new links with high probability giving rise essentially
to a memoryless epidemic dynamics.

The case $\beta=0$ coincides with the memoryless case (ML) and the
dynamics never occurs on the effective static network. Figures ~\ref{fig:fig1}-\ref{fig:fig2}
show that at very small $\beta$, even for the largest considered
value $\left\langle k\right\rangle _{0}=120$, the creation of new
links dominates the dynamics increasing the epidemic threshold towards
the memoryless case. However, we expect that for large enough $\left\langle k\right\rangle _{0}$,
at any $\beta>0$, the dynamics is dominated by memory and infections
spread on the effective static network recovering the ABMF result.
This originates a singular behavior where the limits $\beta\to0$
and $\left\langle k\right\rangle _{0}\to\infty$ do not commute.}

\begin{figure}
\centerline{\includegraphics[width=9cm,height=6cm]{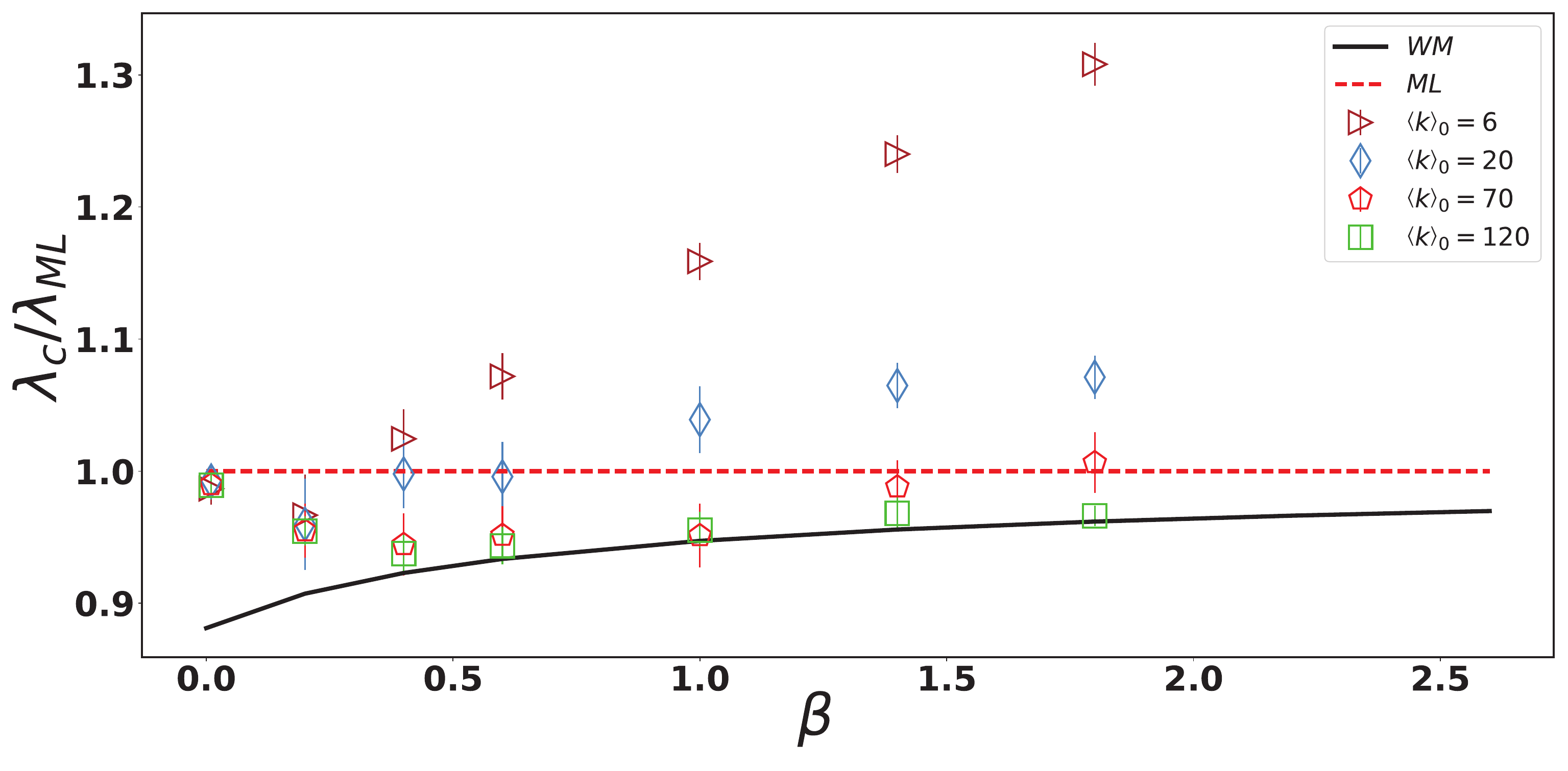}}
\caption{\label{fig:fig3}SIR epidemic process on the activity driven network,
RA. Ratio between the epidemic threshold with memory and the epidemic
threshold of the memoryless case as a function of the reinforcement
parameter $\beta=[0.01,0.2,0.4,0.6,1,1.4,1.8]$. $N=5\cdot10^{4}$,
$\nu=2.4$, $a\in[10^{-2},1]$. The dotted line represents the memoryless
result (ML), the solid line is the analytical prediction obtained
from Eq.~(\ref{eq:cti}) in the memory case (WM). We consider $2\cdot10^{3}$
epidemic realizations for each value of $\lambda$.}
\end{figure}
\begin{figure}
\centerline{\includegraphics[width=9cm,height=6cm]{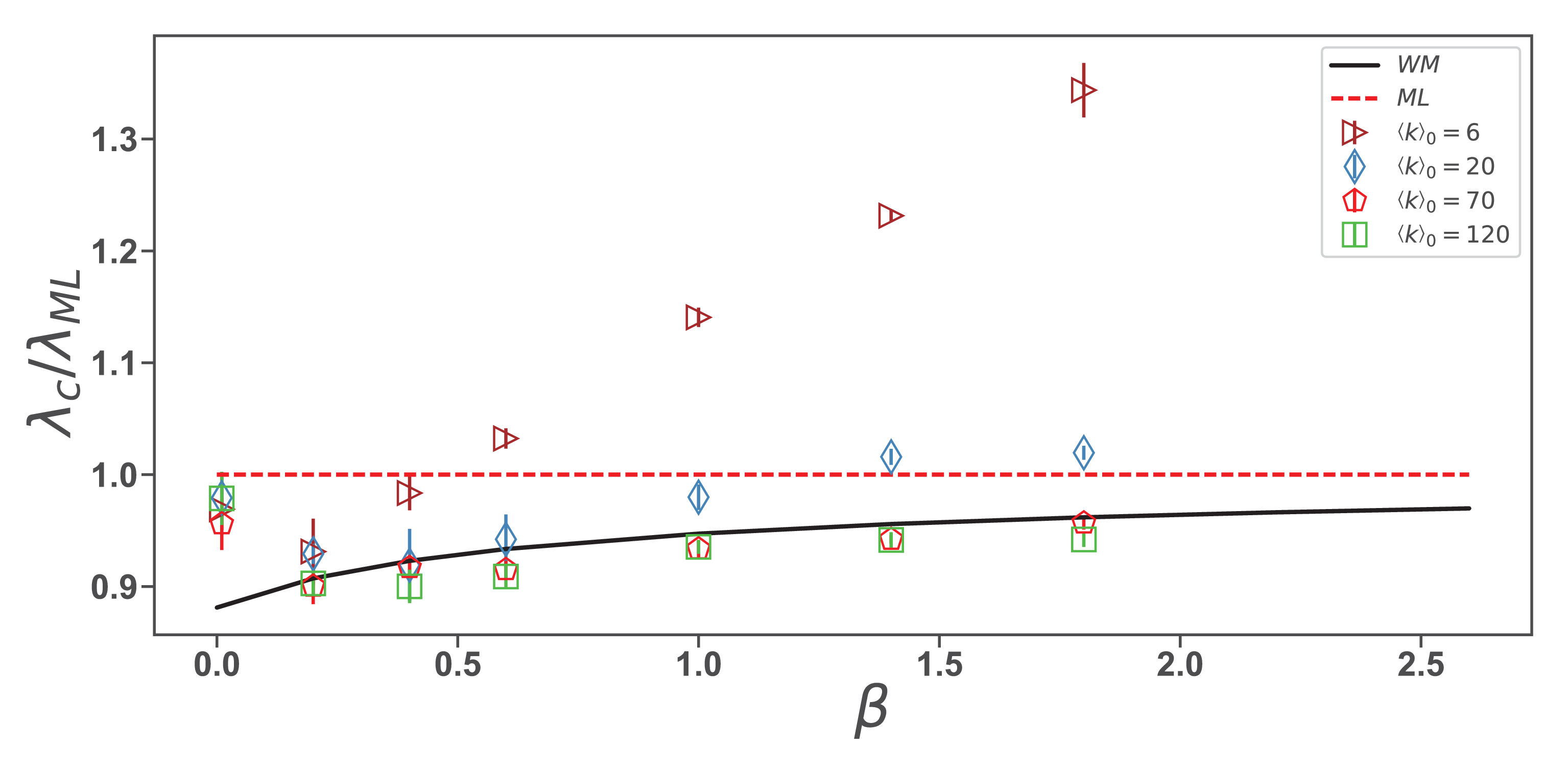}}
\caption{\label{fig:fig4-1}SIR epidemic process on the activity driven network,
MA. Ratio between the epidemic threshold with memory and the epidemic
threshold of the memoryless case as a function of the reinforcement
parameter $\beta=[0.01,0.2,0.4,0.6,1,1.4,1.8]$. $N=5\cdot10^{4}$,
$\nu=2.4$, $a\in[10^{-2},1]$. The dotted line represents the memoryless
result (ML), the solid line is the analytical prediction obtained
from Eq.~(\ref{eq:cti}) in the memory case (WM). We consider $4\cdot10^{3}$
epidemic realizations for each value of $\lambda$.}
\end{figure}

\subsubsection*{SIR model}

The results of simulations of the SIR process are displayed in Fig.
\ref{fig:fig3} and Fig \ref{fig:fig4-1} for the RA and MA case respectively.
The threshold is estimated from the peak of the variability $\Delta=\sqrt{\left\langle N_{R}^{2}\right\rangle -\left\langle N_{R}\right\rangle ^{2}}/\left\langle N_{R}\right\rangle $,
i.e., the standard deviation of the number of recovered nodes $N_{R}$
at the end of the simulation~\citep{doi:10.1063/1.4922153}. As for
SIS, at large $\av{k}_{0}$ dynamical correlations can be neglected
and simulations recover the ABMF result. Simulations clearly show
that now correlations at small $\av{k}_{0}$ inhibit the epidemic
spreading and the critical threshold becomes larger. As in the SIS
model, at small $\beta$ the memory is negligible and the dynamics
is driven by the creation of new links, so that the threshold values
are close to the memoryless case (ML), { even if for any $\beta>0$
we expect for large enough $\left\langle k\right\rangle _{0}$ a convergence
to the ABMF prediction}. On the other hand, for larger $\beta$,
the epidemics evolves on the integrated network, dynamical correlations
become important and the thresholds grow even larger than in the memoryless
case.

\section{Conclusions}

The analytical and numerical results presented in this paper provide
a complete understanding of the interplay between the temporal evolution
of the activity-driven network with memory and the epidemic process
occurring on top of it. The time when the epidemic process begins
has crucial consequences. In the long time limit the reinforcement
mechanism of the topological evolution completely inhibits the formation
of new connections. If the activity-driven epidemic dynamics starts
at this stage, it takes place on a topology which is in practice static.
All nodes have a very large number of connections and this makes mean-field
theory asymptotically exact. The epidemic threshold, which is the
same for SIS and SIR dynamics, is reduced with respect to the memoryless
case, because memory enhances the effect of activity fluctuations.

If instead the epidemic process starts before memory has completely
taken over, interesting model-dependent preasymptotic effects are
observed. The fundamental observation is that at this stage nodes
with large activity tend to interact with their social circles, while
less active nodes still tend to explore the system creating new connections.
The first type of interaction tends to enhance SIS spreading, while
the second tends to suppress it. This leads to positive or negative
corrections to the asymptotic value of the threshold, depending on
the initial conditions and on the reinforcement parameter $\beta$.
In the SIR case instead, since reinfection is not possible, the interaction
within social circles is strongly detrimental for the epidemic propagation,
so that the asymptotic value of the threshold is always reached from
above. Our results allow to fully understand the contrasting effects
of strong ties on SIS and SIR dynamics observed in Ref.~\cite{Sun2015},
and it opens new possibilities in control of epidemic spreading on
temporal networks with high correlations.

Several possible extensions of the model considered here are possible
to make it more realistic, both in terms of the topological evolution
and of the spreading process. Among them probably the most interesting
would be the inclusion of burstiness in agents activity. The combined
effect on activity-driven models of tie reinforcement and non exponentially-distributed
interevent times has been studied in Refs.~\cite{Ubaldi2017a,Ubaldi2017b}.
The inclusion of a spreading dynamics in this framework is a promising
and challenging avenue for future research.

\appendix
{*}

\section{Linear Stability Analysis}
\begin{widetext}
The dynamical process is described by the ABMF equation {[}Eq.~(\ref{eq:cti}){]}
which we rewrite as 
\begin{equation}
\partial_{t}\rho(a)=-\mu\rho(a)+\lambda\left[1-\rho(a)\right]\left[A(a)g(a)\av{\rho(a)}+A(a)\av{g(a)\rho(a)}+g(a)\av{A(a)\rho(a)}+\av{A(a)g(a)\rho(a)}\right],\label{eq:cti-1}
\end{equation}

where for simplicity we have omitted the time dependencies and defined
$A(a)=a/[g(a)+\av{g(a)}]$.

To study the stability of the system linearized around the fixed point
$\rho(a)=0$, we introduce the following functions 
\begin{equation}
\begin{array}{lll}
\rho & = & \av{\rho(a)}\\
x & = & \av{g(a)\rho(a)}\\
y & = & \av{A(a)\rho(a)}\\
z & = & \av{A(a)g(a)\rho(a)}
\end{array}
\end{equation}

Integrating Eq.~(\ref{eq:cti-1}) over $a$ and keeping only linear
terms in $\rho(a)$ we obtain an equation for $\partial_{t}\rho$.
Similarly, multiplying Eq.~(\ref{eq:cti-1}) by $g(a)$ and integrating
over $a$ we get and equation for $\partial_{t}x$. Doing the same
for $y$ and $z$ we obtain a closed system of four equations for
four variables

\begin{equation}
\begin{array}{lll}
\partial_{t}\rho & = & -\mu\rho+\lambda\left[\av{A(a)g(a)}\rho+\av{A(a)}x+\av{g(a)}y+z\right]\\
\partial_{t}x & = & -\mu x+\lambda\left[\av{A(a)g^{2}(a)}\rho+\av{A(a)g(a)}x+\av{g^{2}(a)}y+\av{g(a)}z\right]\\
\partial_{t}y & = & -\mu y+\lambda\left[\av{A^{2}(a)g(a)}\rho+\av{A^{2}(a)}x+\av{A(a)g(a)}y+\av{A(a)}z\right]\\
\partial_{t}z & = & -\mu z+\lambda\left[\av{A^{2}(a)g^{2}(a)}\rho+\av{A^{2}(a)g(a)}x+\av{A(a)g^{2}(a)}y+\av{A(a)g(a)}z\right]
\end{array}
\end{equation}

These equations describe the epidemic near the state $\rho(a)=0$
and the jacobian matrix of this system of equations is 
\begin{equation}
J=\left(\begin{array}{llll}
\lambda\av{Ag}-\mu & \lambda\av{A} & \lambda\av{g} & \lambda\\
\lambda\av{Ag^{2}} & \lambda\av{Ag}-\mu & \lambda\av{g^{2}} & \lambda\av{g}\\
\lambda\av{A^{2}g} & \lambda\av{A^{2}} & \lambda\av{Ag}-\mu & \lambda\av{A}\\
\lambda\av{A^{2}g^{2}} & \lambda\av{A^{2}g} & \lambda\av{Ag^{2}} & \lambda\av{Ag}-\mu
\end{array}\right)
\end{equation}
The state $\rho(a)=0$ is stable provided all eigenvalues of $J$
are negative, hence the epidemic threshold is given by the value $(\lambda/\mu)_{c}$
such that largest eigenvalue of the jacobian matrix is zero. Numerical
evaluation of the matrix $J$ and of its eigenvalues can be obtained,
first by solving numerically Eq. (\ref{Ca}) to get $C(a)$ and $g(a)$
and then calculating the averages defining $J$.
\end{widetext}

\newpage{}


\bibliographystyle{apsrev4-1}
\bibliography{paperS2-fin}

\end{document}